\documentclass[10pt,english,nofootinbib]{revtex4}
\usepackage{lmodern}

\usepackage[T1]{fontenc}
\usepackage[latin9]{inputenc}
\usepackage{amsmath}
\usepackage{amssymb}
\usepackage{esint}

\makeatletter
\@ifundefined{textcolor}{}
{%
 \definecolor{BLACK}{gray}{0}
 \definecolor{WHITE}{gray}{1}
 \definecolor{RED}{rgb}{1,0,0}
 \definecolor{GREEN}{rgb}{0,1,0}
 \definecolor{BLUE}{rgb}{0,0,1}
 \definecolor{CYAN}{cmyk}{1,0,0,0}
 \definecolor{MAGENTA}{cmyk}{0,1,0,0}
 \definecolor{YELLOW}{cmyk}{0,0,1,0}
 }

\@ifundefined{definecolor}{\@ifundefined{definecolor}{\@ifundefined{definecolor}{\@ifundefined{definecolor}
 {\usepackage{color}}{}
}{}}{}}{}\@ifundefined{definecolor}{\@ifundefined{definecolor}{\@ifundefined{definecolor}{\@ifundefined{definecolor}{\@ifundefined{definecolor}
 {\usepackage{color}}{}
}{}}{}}{}}{}\@ifundefined{definecolor}{\@ifundefined{definecolor}{\@ifundefined{definecolor}{\@ifundefined{definecolor}{\@ifundefined{definecolor}{\@ifundefined{definecolor}
 {\usepackage{color}}{}
}{}}{}}{}}{}}{}\@ifundefined{definecolor}{\@ifundefined{definecolor}{\@ifundefined{definecolor}{\@ifundefined{definecolor}{\@ifundefined{definecolor}{\@ifundefined{definecolor}{\@ifundefined{definecolor}
 {\usepackage{color}}{}
}{}}{}}{}}{}}{}}{}\usepackage{latexsym}\usepackage{bm}

\makeatother

\makeatother

\makeatother

\usepackage{babel}

\makeatother

\usepackage{babel}

\makeatother

\usepackage{babel}

\makeatother

\usepackage{babel}

\makeatother

\usepackage{babel}

\makeatother

\usepackage{babel}

\makeatother

\usepackage{babel}

\begin{document}
\hfill{}{{\small ~BRX-TH-632,~~~CALT 68-281,~~~MIFPA-11-04~}~}

\title{Critical Points of $D$-Dimensional Extended Gravities}

\author{S. Deser}

\email{deser@brandeis.edu}

\affiliation{Physics Department, Brandeis University, Waltham, MA 02454, USA}

\affiliation{Lauritsen Laboratory, California Institute of Technology, Pasadena,
CA 91125, USA}

\author{Haishan Liu}

\email{liu@zimp.zju.edu.cn}

\affiliation{Zheijiang Institute of Modern Physics, Department of Physics, Zheijiang
University, Hangzhou 310027, China}

\author{H. Lü}

\email{mrhonglu@gmail.com}

\affiliation{China Economics and Management Academy, Central University of Finance
and Economics, Beijing 100081}

\affiliation{Institute for Advanced Study, Shenzhen University, Nanhai Ave 3688,
Shenzhen 518060, China}

\author{C.N. Pope}

\email{pope@physics.tamu.edu}

\affiliation{George P. \& Cynthia Woods Mitchell Institute for Fundamental Physics
and Astronomy, Texas A\&M University, College Station, TX 77843, USA}

\affiliation{DAMTP, Centre for Mathematical Sciences, Cambridge University, Wilberforce
Road, Cambridge CB3 OWA, UK}

\author{Tahsin Ça\u{g}r\i{} \c{S}i\c{s}man}

\email{sisman@metu.edu.tr}

\affiliation{Department of Physics, Middle East Technical University, 06531, Ankara,
Turkey}

\author{Bayram Tekin}

\email{btekin@metu.edu.tr}

\affiliation{Department of Physics, Middle East Technical University, 06531, Ankara,
Turkey}

\date{\today}
\begin{abstract}
We study the parameter space of $D$-dimensional cosmological Einstein
gravity together with quadratic curvature terms. In $D>4$ there are
in general two distinct (anti)-de Sitter vacua. We show that for appropriate
choice of the parameters there exists a critical point for one of
the vacua, for which there are only massless tensor, but neither massive
tensor nor scalar, gravitons. At criticality, the linearized excitations
have vanishing energy (as do black hole solutions). A further restriction
of the parameters gives a one-parameter cosmological Einstein plus Weyl$^2$ 
model with a unique vacuum, whose $\Lambda$ is determined.
\end{abstract}
\maketitle

\subsubsection{Introduction}

Adding new types of terms--and their associated parameters--to an
action can lead, in the extended parameter space, to solutions with
qualitatively different properties than in the component pieces. In
the gravitational models we consider here, such phenomena have already
been found, even at the linearized level. Historical examples include
multiple maximally-symmetric vacua, one flat and the other (A)dS in
Einstein-Hilbert gravity with quadratic curvature terms in dimensions
$D\ne4$ \cite{boudes}; {}``partial masslessness'' \cite{deswal},
in which degrees of freedom are lost and gauge invariance {}``emerges''
for massive higher spins $s>1$ in AdS backgrounds with suitable mass$^{2}/\Lambda$
ratios; and the tuning of masses in the {}``new massive gravity''
(NMG) in $D=3$ \cite{BHT,lisu}. Another example where the tuning
of parameters leads to new effects is the so-called {}``chiral gravity''
\cite{LiSongStrominger}, which is a special case of topologically
massive gravity (TMG) in AdS \cite{DeserJackiwTempleton}. Most recently
\cite{LuPope}, still wider critical effects were found in $D=4$
systems combining the Einstein-Hilbert action, a cosmological term,
and terms quadratic in the curvature. The latter work motivates our
extension to generic dimensions, where the interplay between the multiple
parameters will provide novel effects, some relying on the explicit
cosmological term. For example, as we shall see, multiple vacua and
{}``tuned'' masslessness will now interact. We will find these critical
models by studying their conserved charges, especially the energy,
of black holes, as well as their linearized excitation spectrum.

\subsubsection{Field equations }

\noindent The most general quadratic gravity model is \begin{equation}
I=\int d^{D}x\,\sqrt{-g}\left[\frac{1}{\kappa}\left(R-2\Lambda_{0}\right)+\alpha R^{2}+\beta R^{\mu\nu}R_{\mu\nu}+\gamma\left(R^{\mu\nu\rho\sigma}R_{\mu\nu\rho\sigma}-4R^{\mu\nu}R_{\mu\nu}+R^{2}\right)\right].\label{eq:Quadratic_action}\end{equation}
 We follow the notation and conventions of \cite{DeserTekin,GulluTekin},
where relevant details of field equations, curvature linearizations,
and conserved charges may also be found. For generic values of the
parameters, there are two distinct (A)dS vacua%
\footnote{Depending upon the choice of parameters, and assuming $\Lambda_{0}<0$,
there can be either two AdS vacua or else one AdS and one dS. If $\Lambda_{0}$
is positive, there will be either two dS vacua or one dS and one AdS.
If $\Lambda_{0}=0$ there will be one Minkowski vacuum and either
an AdS or a dS vacuum.%
}, with metrics satisfying $R_{\mu\nu}=\frac{2\Lambda}{D-2}\, g_{\mu\nu}$,
where $\Lambda$ is determined by the quadratic equation \begin{equation}
\frac{\Lambda-\Lambda_{0}}{2\kappa}+f\Lambda^{2}=0,\qquad f\equiv\left(D\alpha+\beta\right)\frac{\left(D-4\right)}{\left(D-2\right)^{2}}+\gamma\frac{\left(D-3\right)\left(D-4\right)}{\left(D-1\right)\left(D-2\right)}.\label{quadratic}\end{equation}
 This phenomenon was first observed \cite{boudes} in Einstein-Lovelock
gravity ($\alpha=\beta=0$, $\Lambda_{0}=0$), where Minkowski and
(A)dS vacua coexist. The source-free linearized equation of motion
for the metric fluctuations around an (A)dS vacuum, $h_{\mu\nu}\equiv g_{\mu\nu}-\bar{g}_{\mu\nu}$,
becomes \begin{equation}
c\,\mathcal{G}_{\mu\nu}^{L}+\left(2\alpha+\beta\right)\left(\bar{g}_{\mu\nu}\bar{\square}-\bar{\nabla}_{\mu}\bar{\nabla}_{\nu}+\frac{2\Lambda}{D-2}\bar{g}_{\mu\nu}\right)R^{L}+\beta\left(\bar{\square}\mathcal{G}_{\mu\nu}^{L}-\frac{2\Lambda}{D-1}\bar{g}_{\mu\nu}R^{L}\right)=0,\label{eq:Linearized_eom}\end{equation}
 where $c$ is given by \[
c\equiv\frac{1}{\kappa}+\frac{4\Lambda D}{D-2}\alpha+\frac{4\Lambda}{D-1}\beta+\frac{4\Lambda\left(D-3\right)\left(D-4\right)}{\left(D-1\right)\left(D-2\right)}\gamma,\]
 and the linearization of the Einstein tensor ${\mathcal{G}}_{\mu\nu}\equiv 
R_{\mu\nu}-\frac12 R g_{\mu\nu} +\Lambda g_{\mu\nu}$ (defined using the
effective $\Lambda$ of the system's vacuum state(s), not the
bare parameter $\Lambda_0$ in the action) is
given by
 \[
\mathcal{G}_{\mu\nu}^{L}=R_{\mu\nu}^{L}-\frac{1}{2}\bar{g}_{\mu\nu}R^{L}-\frac{2\Lambda}{D-2}h_{\mu\nu}.\]
 Here, the linearized Ricci tensor $R_{\mu\nu}^{L}$ and scalar
curvature $R^{L}=\left(g^{\mu\nu}R_{\mu\nu}\right)^{L}$ are given
by \[
R_{\mu\nu}^{L}=\frac{1}{2}\left(\bar{\nabla}^{\sigma}\bar{\nabla}_{\mu}h_{\nu\sigma}+\bar{\nabla}^{\sigma}\bar{\nabla}_{\nu}h_{\mu\sigma}-\bar{\square}h_{\mu\nu}-\bar{\nabla}_{\mu}\bar{\nabla}_{\nu}h\right),\qquad R^{L}=-\bar{\square}h+\bar{\nabla}^{\sigma}\bar{\nabla}^{\mu}h_{\sigma\mu}-\frac{2\Lambda}{D-2}h.\]

Taking the trace of (\ref{eq:Linearized_eom}) gives \begin{equation}
\left[\left(4\alpha\left(D-1\right)+D\beta\right)\bar{\square}-\left(D-2\right)\left(\frac{1}{\kappa}+4f\Lambda\right)\right]R^{L}=0.\label{eq:Trace_lin_eom}\end{equation}
 We see that the D'Alembertian operator is removed if $\alpha$ and
$\beta$ are chosen so that \begin{equation}
4\alpha\left(D-1\right)+D\beta=0,\label{albe}\end{equation}
 and so $R^{L}$ is constrained to vanish, provided that \begin{equation}
\frac{1}{\kappa}+4f\Lambda\ne0.\label{exception}\end{equation}
 (See \cite{GulluTekin}, and the discussion below.) As in \cite{LuPope},
we shall choose the gauge $\bar{\nabla}^{\mu}h_{\mu\nu}=\bar{\nabla}_{\nu}h$,
which then leads to $R^{L}=-\frac{2\Lambda}{D-2}h$. Imposing the
condition (\ref{albe}), which eliminates the scalar mode (see equation
(30) of \cite{GulluTekin}), one has $h=0$ from (\ref{eq:Trace_lin_eom}),
and hence $h_{\mu\nu}$ satisfies the transverse and traceless conditions
\begin{equation}
\bar\nabla^{\mu}h_{\mu\nu}=0,\qquad h=0.\label{TTgauge}\end{equation}
 The linearized Ricci tensor $R_{\mu\nu}^{L}$ and the linearized
Einstein tensor $\mathcal{G}_{\mu\nu}^{L}$ in this transverse-traceless
gauge become \[
R_{\mu\nu}^{L}=\frac{2D\Lambda}{\left(D-1\right)\left(D-2\right)}h_{\mu\nu}-\frac{1}{2}\bar{\square}h_{\mu\nu},\qquad\mathcal{G}_{\mu\nu}^{L}=\frac{2\Lambda}{\left(D-1\right)\left(D-2\right)}h_{\mu\nu}-\frac{1}{2}\bar{\square}h_{\mu\nu}.\]
 The equations of motion simplify to\begin{equation}
-\frac{\beta}{2}\left(\bar{\square}-\frac{4\Lambda}{\left(D-1\right)\left(D-2\right)}-M^{2}\right)\left(\bar{\square}-\frac{4\Lambda}{\left(D-1\right)\left(D-2\right)}\right)h_{\mu\nu}=0,\label{4eom}\end{equation}
 where \begin{equation}
M^{2}\equiv-\frac{1}{\beta}\left(c+\frac{4\Lambda\beta}{\left(D-1\right)\left(D-2\right)}\right).\label{eq:Mass_of_spin-2_excitation}\end{equation}
 The massless and massive modes satisfy \begin{equation}
\left(\bar{\square}-\frac{4\Lambda}{\left(D-1\right)\left(D-2\right)}\right)h_{\mu\nu}^{\left(m\right)}=0,\qquad\left(\bar{\square}-\frac{4\Lambda}{\left(D-1\right)\left(D-2\right)}-M^{2}\right)h_{\mu\nu}^{\left(M\right)}=0,\label{modes}\end{equation}
 respectively. Stability requires $M^{2}\ge0$, and the case where
$M^{2}=0$, i.e. \begin{equation}
c+\frac{4\Lambda\beta}{\left(D-1\right)\left(D-2\right)}=0,\label{eq:critical}\end{equation}
 defines the critical point.

\subsubsection{Energy }

\noindent We now show that at the critical point, the mass and angular
momenta of all asymptotically Kerr-AdS and Schwarzschild-AdS black
holes vanish.%
\footnote{\noindent Actually, unlike for cosmological Einstein theory \cite{Gibbons},
explicit Kerr-AdS type solutions for these models are as yet unknown.
Such solutions presumably exist, and they will approach the standard
Kerr-AdS metrics at large distance where the effects of the higher-order
curvature terms become negligible.%
} We can calculate the conserved charges of any such asymptotically
Kerr-AdS space, as discussed in \cite{Kanik}. Letting $\bar{\xi}_{\mu}$
be a Killing vector of AdS, the conserved charges associated with
this Killing vector will be given by \cite{DeserTekin}\begin{align}
Q^{\mu}(\bar{\xi})= & \left(c+\frac{4\Lambda\beta}{\left(D-1\right)\left(D-2\right)}\right)\int_{\mathcal{M}}d^{D-1}x\,\sqrt{-\bar{g}}\bar{\xi_{\nu}}\mathcal{G}_{L}^{\mu\nu}\nonumber \\
 & +(2\alpha+\beta)\int_{\partial\mathcal{M}}dS_{i}\sqrt{-g}\left\{ \bar{\xi}^{\mu}\bar{\nabla}^{i}R_{L}+R_{L}\bar{\nabla}^{\mu}\,\bar{\xi}^{i}-\bar{\xi}^{i}\bar{\nabla}^{\mu}R_{L}\right\} \nonumber \\
 & +\beta\int_{\partial\mathcal{M}}dS_{i}\sqrt{-g}\left\{ \bar{\xi}_{\nu}\bar{\nabla}^{i}\mathcal{G}_{L}^{\mu\nu}-\bar{\xi}_{\nu}\bar{\nabla}^{\mu}\mathcal{G}_{L}^{i\nu}-\mathcal{G}_{L}^{\mu\nu}\bar{\nabla}^{i}\bar{\xi}_{\nu}+\mathcal{G}_{L}^{i\nu}\bar{\nabla}^{\mu}\bar{\xi}_{\nu}\right\} .\label{fullcharge}\end{align}
 Here, $\mathcal{M}$ is a spacelike hypersurface in the asymptotically
AdS space, and $\partial\mathcal{M}$ denotes its boundary. The first
integral can also be written as a boundary term, see \cite{Abbott,DeserTekin}.
In asymptotically AdS spaces, only the first term survives. (We recall
that the charges in (\ref{fullcharge}) are measured at infinity;
their detailed form as volume integrals is given by the spatial integral
of the field equations' nonlinear terms.) At the critical point then,
the charges, and in particular the energy ($Q^{0}$) vanishes. For
example, the energy of the Schwarzschild-AdS solution, with asymptotics
$h_{00}=h^{rr}=\left(\frac{r_{0}}{r}\right)^{D-3}$, is \cite{DeserTekin}
\begin{equation}
E_{BH}=\left(c+\frac{4\Lambda\beta}{\left(D-1\right)\left(D-2\right)}\right)\frac{\left(D-2\right)}{2}\Omega_{D-2}r_{0}^{D-3},\label{eq:Energy}\end{equation}
 where $\Omega_{D-2}$ is the solid angle on the $D-2$ sphere. Specifically,
in four dimensions, $\kappa=8\pi G_{N}$ and $r_{0}=2G_{N}m$, and
then $E_{BH}=\left[1+16\pi G_{N}\left(4\alpha+\beta\right)\right]m$.
At the critical point (\ref{eq:critical}), we see that in all dimensions,
$E_{BH}=0$.

Note that in $D=3$, the critical theory reduces to the NMG at the
{}``Proca'' point. Namely, $8\alpha+3\beta=0$ gives the NMG theory,
and the criticality condition $M^{2}=0$ gives $\Lambda_{0}=\frac{3}{\kappa\beta}$
which is exactly the point where the linearized NMG can be written
explicitly as a unitary massive spin-1 theory with mass squared $-\frac{8}{\kappa\beta}$
(see \cite{lisu}, and Sec. V.A in \cite{BHT}). (For AdS and with
$\beta>0$, one should set $\kappa<0$ since $\Lambda=\frac{2}{\kappa\beta}$.)

\subsubsection{Linear excitations}

\noindent Let us now consider the energy of the excitations, expressed
as explicit volume integrals, by constructing the Hamiltonian for
the theory, as in \cite{LiSongStrominger,LuPope}. The quadratic Lagrangian
for the metric perturbations of (\ref{eq:Quadratic_action}) around
an AdS vacuum is given by simply multiplying the left-hand side of
(\ref{eq:Linearized_eom}) by $-\frac{1}{2}h^{\mu\nu}$; hence it
vanishes on-shell. With the parameter choice (\ref{albe}), and in
the transverse-traceless gauge (\ref{TTgauge}), the quadratic action
therefore takes the form \begin{equation}
I_{2}=-\frac{1}{2}\int d^{D}x\,\sqrt{-\bar{g}}\left[-\frac{\beta}{2}\bar{\square}h^{\mu\nu}\bar{\square}h_{\mu\nu}-\frac{1}{2}\left(\frac{4\Lambda\beta}{\left(D-1\right)\left(D-2\right)}-c\right)\bar{\nabla}^{\rho}h^{\mu\nu}\bar{\nabla}_{\rho}h_{\mu\nu}+\frac{2\Lambda c}{\left(D-1\right)\left(D-2\right)}h^{\mu\nu}h_{\mu\nu}\right].\label{eq:Oh2_action_in_TT_gauge}\end{equation}
 The canonical momenta, using the Ostrogradsky formalism for higher-order
Lagrangians, are defined and computed as follows: \begin{align*}
\Pi_{\left(1\right)}^{\mu\nu} & \equiv\frac{\delta\mathcal{L}_{2}}{\delta\dot{h}_{\mu\nu}}-\bar{\nabla}_{0}\left(\frac{\delta\mathcal{L}_{2}}{\delta\left[\frac{\partial}{\partial t}\left(\bar{\nabla}_{0}h_{\mu\nu}\right)\right]}\right)=-\frac{\sqrt{-\bar{g}}}{2}\bar{\nabla}^{0}\left[-\left(\frac{4\Lambda\beta}{\left(D-1\right)\left(D-2\right)}-c\right)h^{\mu\nu}+\beta\bar{\square}h^{\mu\nu}\right],\\
\Pi_{\left(2\right)}^{\mu\nu} & \equiv\frac{\delta\mathcal{L}_{2}}{\delta\left[\frac{\partial}{\partial t}\left(\bar{\nabla}_{0}h_{\mu\nu}\right)\right]}=\frac{\sqrt{-\bar{g}}}{2}\beta\bar{g}^{00}\bar{\square}h^{\mu\nu}.\end{align*}
 The Hamiltonian can be written as \begin{equation}
H\equiv\int d^{D-1}x\,\left[\Pi_{\left(1\right)}^{\mu\nu}\dot{h}_{\mu\nu}+\Pi_{\left(2\right)}^{\mu\nu}\frac{\partial}{\partial t}\left(\bar{\nabla}_{0}h_{\mu\nu}\right)\right]-\int\sqrt{-g}\mathcal{L}_{2}d^{D-1}x,\label{Hamiltonian}\end{equation}
 where we use for the implicit AdS background metric the time-independent
form \[
d\bar{s}^{2}=\frac{(D-1)(D-2)}{2(-\Lambda)}\,\Big[-\cosh^{2}\rho dt^{2}+d\rho^{2}+\sinh^{2}\rho d\Omega_{D-2}^{2}\Big].\]
 Substituting the canonical momenta into (\ref{Hamiltonian}) gives
the Hamiltonian for the quadratic fluctuations. The energies of the
various (linearized) excitations could be obtained directly from (\ref{Hamiltonian}),
but a more convenient starting point is to note that the Hamiltonian
$H$ is time-independent, since the Lagrangian has no explicit time
dependence. We are therefore free to write $H$ as its time average,
$H=\langle H\rangle=T^{-1}\int_{0}^{T}dtH$. The advantage of doing
this is that we can then perform integrations by parts for time derivatives.
This allows us to manipulate (\ref{Hamiltonian}) into a more convenient
form. The Lagrangian term $\int\sqrt{-g}\mathcal{L}_{2}d^{D-1}x$,
being proportional to the equation of motion (\ref{4eom}), does not
contribute to the on-shell energies. Substituting the expressions
(\ref{modes}) for the massless and the massive modes into the time
average of the Hamiltonian (\ref{Hamiltonian}), we find, after an
integration by parts, that their energies are given by \begin{align}
E_{m} & =-\frac{1}{2T}\left(c+\frac{4\Lambda\beta}{\left(D-1\right)\left(D-2\right)}\right)\int d^{D}x\,\sqrt{-\bar{g}}\left(\dot{h}_{\mu\nu}^{\left(m\right)}\bar{\nabla}^{0}h_{\left(m\right)}^{\mu\nu}\right),\\
E_{M} & =\frac{1}{2T}\left(c+\frac{4\Lambda\beta}{\left(D-1\right)\left(D-2\right)}\right)\int d^{D}x\,\sqrt{-\bar{g}}\left(\dot{h}_{\mu\nu}^{\left(M\right)}\bar{\nabla}^{0}h_{\left(M\right)}^{\mu\nu}\right),\end{align}
 where the time integrations are understood to be over the interval
$T$. The excitation energies $E_{m}$ and $E_{M}$ have opposite
signs, as expected from their fourth-order origin \cite{stelle},
but at the critical point (\ref{eq:critical}), they vanish. The integral
itself for the massless modes is known to be negative, since in pure
Einstein gravity for which $\alpha=\beta=\gamma=0$, the energies
$E_{m}$ are known to be positive. The integral for the massive modes
is also expected to be negative, away from $M^{2}=0$, at least for
small mass.

We started with a theory with four parameters, namely $\Lambda_{0}$,
$\alpha$, $\beta$ and $\gamma$ (we are not counting $\kappa$,
since it can be scaled out). The critical point is defined by the
two conditions (\ref{albe}) and (\ref{eq:critical}), implying that
any two of the four parameters can be eliminated. Note, however, that
we must also require that (\ref{exception}) hold at the critical
point, since otherwise, $R_{L}$, a gauge invariant object, would
be left undetermined in the theory.

As we already remarked, there are in general two distinct AdS vacua
in the theory with given values for the four parameters $\Lambda_{0}$,
$\alpha$, $\beta$ and $\gamma$, corresponding to the two roots
of the quadratic equation (\ref{quadratic}). The specialisation to
the critical case that we have been discussing turns the massive spin-2
modes massless for just one of these vacua. In the other vacuum, the
spectrum still contains massive as well as massless spin-2 modes,
with excitation energies of opposite signs. Whether it is the massive
spin-2 modes or the massless spin-2 modes that have the negative energy
depends upon the detailed choice of parameters. Note that if the parameters
are chosen such that the massless spin-2 modes have positive energy,
then it will also be the case that black holes in the non-critical
vacuum will have positive mass.

For some purposes, in order to avoid solving the quadratic equation
(\ref{quadratic}) for $\Lambda$, it is convenient instead to view
it as a linear equation for the parameters of the theory expressed
in terms of the critical value of $\Lambda$. It seems to be most
convenient to take $\alpha$ and $\gamma$ as the free parameters
for $D>4$. From (\ref{albe}), one determines $\beta$ in terms of
$\alpha$. Then, from (\ref{eq:critical}), one obtains the unique
critical vacuum with a cosmological constant \begin{equation}
\Lambda_{\text{crit}}=-\frac{D\left(D-1\right)\left(D-2\right)}{4\kappa\left[\left(D-1\right)\left(D-2\right)^{2}\alpha+D\left(D-3\right)\left(D-4\right)\gamma\right]}.\label{eq:Critic_vac}\end{equation}
 To determine the corresponding critical $\Lambda_{0}$, we can use
(\ref{quadratic}) to get\begin{equation}
\Lambda_{0}=-\frac{D^{2}\left(D-1\right)\left(D-2\right)\left[\left(D-1\right)\left(D-2\right)\alpha+\left(D-3\right)\left(D-4\right)\gamma\right]}{8\kappa\left[\left(D-1\right)\left(D-2\right)^{2}\alpha+D\left(D-3\right)\left(D-4\right)\gamma\right]^{2}}.\label{eq:Lambda0}\end{equation}
 The other (non-critical) vacuum has the cosmological constant \begin{equation}
\Lambda_{\text{noncrit}}=\frac{D[(D-1)(D-2)\alpha+(D-3)(D-4)\gamma]\,\Lambda_{\text{crit}}}{(D-4)[(D-1)(D-2)\alpha+D(D-3)\gamma]}.\label{eq:Noncritic_vac}\end{equation}
 As a result, the two-parameter critical theory is given by the action
\begin{equation}
I=\int d^{D}x\,\sqrt{-g}\left[\frac{1}{\kappa}\left(R-2\Lambda_{0}\right)+\alpha R^{2}-\frac{4\left(D-1\right)}{D}\alpha R^{\mu\nu}R_{\mu\nu}+\gamma\left(R^{\mu\nu\rho\sigma}R_{\mu\nu\rho\sigma}-4R^{\mu\nu}R_{\mu\nu}+R^{2}\right)\right],\end{equation}
 where the bare cosmological constant is given by (\ref{eq:Lambda0}).
The theory has a critical vacuum (\ref{eq:Critic_vac}), while there
is a noncritical vacuum (\ref{eq:Noncritic_vac}).

Finally, one can overcome the two-vacuum problem, that one of them
necessarily allows negative-energy excitations, using one last allowed
(consistent as we shall see) reduction, to a single parameter. We
require $f=0$ in (\ref{quadratic}), whose only solution is thereby
$\Lambda=\Lambda_{0}$. Our four parameters ($\alpha$, $\beta$,
$\gamma$, $\Lambda_{0}$), are now constrained by the three conditions
(\ref{quadratic}, \ref{eq:critical}), plus scalar mode suppression
(\ref{albe}), to obey \begin{equation}
\beta=-\frac{4\alpha\left(D-1\right)}{D},\qquad\Lambda=\Lambda_{0}=-\frac{D}{8\kappa\alpha},\qquad\alpha=-\frac{\gamma D\left(D-3\right)}{\left(D-1\right)\left(D-2\right)}.\end{equation}
 Note that $\Lambda_{0}$ is also fixed in terms of the remaining
parameter. The general action (\ref{eq:Quadratic_action}) now reduces
to Einstein-Weyl form with a single arbitrary constant $\gamma$,
\begin{equation}
I=\int d^{D}x\,\sqrt{-g}\left[\frac{1}{\kappa}\left(R-2\Lambda_{0}\right)+\gamma C^{\mu\nu\rho\sigma}C_{\mu\nu\rho\sigma}\right],\qquad\Lambda=\Lambda_{0}=\frac{\left(D-1\right)\left(D-2\right)}{8\kappa\gamma\left(D-3\right)}.\label{weylaction}\end{equation}
 Here, $C_{\mu\nu\rho\sigma}$ is the Weyl tensor, for which \[
C^{\mu\nu\rho\sigma}C_{\mu\nu\rho\sigma}=R^{\mu\nu\rho\sigma}R_{\mu\nu\rho\sigma}-\frac{4}{D-2}R^{\mu\nu}R_{\mu\nu}+\frac{2}{\left(D-1\right)\left(D-2\right)}R^{2}.\]
 Physically, it is clear a priori that only the combination (\ref{weylaction})
has a single vacuum: the Weyl tensor, being insensitive to conformally
related metrics, such as different (A)dS or flat ones, does not contribute
directly to (\ref{quadratic}). We also emphasize that while condition
(\ref{exception}) is automatically satisfied by taking $f=0$, attempting
instead to force the two roots of (\ref{quadratic}) to be equal would
violate it.

\subsubsection{Conclusions}

\noindent We have constructed a two-parameter theory of gravity in
$D$ dimensions, which admits a critical AdS vacuum in which there
are only massless spin-2 modes. For $D=3$, the critical theory reduces
to the NMG at the Proca point, whilst for $D=4$, it reduces to the
critical theory constructed recently in \cite{LuPope}. In four dimensions,
once the condition (\ref{albe}) for the elimination of the scalar
mode is imposed, the theory has a unique AdS vacuum solution. For
dimensions $D\ne4$, by contrast, the theory in general has two distinct
AdS vacua, which cannot simultaneously be rendered critical for any
choice of parameters. For $D\ge5$, a further specialisation of parameters
can be made, leading to a one-parameter critical theory with a unique
AdS vacuum. The surprising confluence of so many unexpected properties
at our models' critical points may repay further investigation.

\section{Acknowledgments}

S.D. is supported by NSF PHY 07-57190 and DOE DE-FG02-164 92ER40701
grants. H. Liu is supported in part by the National Science Foundation
of China (10425525,10875103), National Basic Research Program of China
(2010CB833000) and Zhejiang University Group Funding (2009QNA3015).
The research of C.N.P. is supported in part by DOE grant DE-FG03-95ER40917.
The work of T.C.S. and B.T. is supported by the TÜB\.{I}TAK Grant
No. 110T339, and METU Grant BAP-07-02-2010-00-02. H. Lü is grateful
to the Mitchell Institute for Fundamental Physics and Astronomy at
Texas A\&M University for hospitality during the course of this work.

\end{document}